# AI Algorithm for the Generation of Three-Dimensional Accessibility Ramps in Grasshopper / Rhinoceros 7

Antonio Sitong Li
UWC Changshu China
ali21@uwcchina.org

Leila Wen Xuan Yi
UWC Changshu China
lyi21@uwcchina.org

Brandon Yeo
UWC Changshu China
byeo@uwcchina.org

*Abstract* - Often overlooked as a component of urban development, accessibility infrastructure is undeniably crucial in daily life. Accessibility ramps are one of the most common types of accessibility infrastructure, and serve to benefit not only people with mobile impairments but also able-bodied third parties. While the necessity of accessibility ramps is acknowledged, actual implementation fails in light of the limits of manpower required for the design stage.

In response, we present an algorithm capable of the automatic generation of a feasible accessibility ramp based on a 3D model of the relevant environment. Through the manual specification of initial and terminal points within a 3D model, the algorithm uses AI search algorithms to determine the optimal pathway connecting these points. Essential components in devising a wheelchair-accessible ramp are encoded within the process, as evaluated by the algorithm, including but not limited to elevation differentials, spatial constraints, and gradient specifications. From this, the algorithm then generates the pathway to be expanded into a full-scale, useable model of a ramp, which then can be easily exported and transformed through inter-software exchanges. Though some human input is still required following the generation stage, the minimising of human resources provides significant boosts of efficiency in the design process thus lowering the threshold for the incorporation of accessibility features in future urban design.

**Keywords**: path generation, terrain traversal, A* algorithm, parametric design, 3D modelling, software

**I. INTRODUCTION**

Accessibility refers to an individual's ability to be mobile and partake in the enjoyment of services, receiving of information, and execution of activities in a dignified, independent, and safe manner [1]. In reference to the World Health Organisation's 2011 report on disability [2], an estimated 15% of the population worldwide live with a form of disability, with this percentage of the population increasing in consideration of an ageing society and population growth. An international call for greater equity and accessibility has long been set in motion [3], focusing on hammering the technological advancements of the recent decades to tackle this deficit.

Infrastructure, however, continues to maintain its grounds as one of the most powerful tools for empowering and reintegrating the disabled populace into spaces meant to pivot societal and economic development [4]. Even on a day-to-day basis, accessibility infrastructure is an essential means for the long-standing chasm between individuals with mobile-impairments and the social community they belong to. With the wide-spread implementation of accessibility infrastructure, said individuals are able to regain their access to certain spaces and repossess a greater degree of independence [5].

Accessibility ramps are one of the most popular and commonly implemented pieces of accessibility infrastructures, with many pieces of relevant policies and literature published on the standards of public-use ramps [6]. Considering the degree of third-person benefit from construction and implementation of accessibility ramps in public spaces, detailed technical requirements to reinsert the central purpose of said ramps were published. To prevent urban architecture from neglecting the purpose of the ramps (to increase accessibility), hallmark publications such as the ADA (Americans with Disabilities Act) [7, 8] enforce strict regulations on elements of ramp specification. These specifications include ramp slope, width, landing spaces, handrail requirements, cross slopes, edge protection, and so on and so forth [7]. Challenges arise from the actual implementation of ramp designs in buildings, with the often-underestimated requirements for specialised knowledge and invested time being its detriment [9, 10].

Seamless integration of ramps into building layouts mandates careful consideration of structures in the existing environment, as well as careful planning. Involved designers and architects [9] are required to allocate copious lengths of time to the initial design process for the sake of cohesive ramp integration and the prevention of compromising the building's design, aesthetics, and functionality [10]. As such, a sufficient level of expertise is required in almost every quality ramp-integration design project. Understandings of relevant accessibility guidelines and user needs, local building regulations and codes, as well as field expertise, are all key to the provision of a genuinely accessible experience [11, 12]. For the successful execution of such projects, collaboration between a team of designers, architects, engineers, accessibility consultants, as well as construction personnel would all be required to work in tandem. The costs [13] of such conditions would, then, act as highly restrictive prerequisites for the wide-spread implementation of accessibility ramps in urban infrastructure.

Examining the rapid urbanisation of modern China, a period of economic reform led to a large-scale migration of an estimate of more than 600 million from rural China to the main cities [14]. As such, this movement was reflected in a growth in urbanisation rate from 17.9% previously to a tumultuous 64.72% following [14, 15]. As expected, a miscellany of socio-economic repercussions soon followed, ranging from environmental damage to urban vulnerability in most major Chinese cities. To deal with these newly-arising issues, a process of significant, planned urban reconstruction was prescribed to these degrading spaces [16], upgrading the structural and industrial systems within these areas, as well as promoting a gradual evolution of all hotspot cities in general.

Released in conjunction with the plans for urban reconstruction [17], the Chinese government has placed a newly significant focus on accessibility features in urban redesign [17, 18]. Placing emphasis on the smooth integration of such infrastructure in all aspects of building design, accessibility for all individuals of all levels of ability is regarded as one of the core considerations for the redesign. This new promise of increased accessibility as a focus, rather than an addition [19, 20] to urban infrastructure signifies the possibility of a more equitable and more accessibly-constructed future [21].

### A. Rationale

Parametric design refers to a design process using computational capabilities [22] involving deduction, induction, and abstraction, which is inspired from algorithmic thinking [22, 23], while enforcing certain parameters and rules to constrain them. Associative geometry and topological relationships are used for establishing dependencies among design elements and are also reinvented and applied to contents such as the BIM paradigm [24]. To bring the focus to parametric architecture specifically, the obsession with relations between form parameters can be observed once more. Real-world examples of the implementation of parametric architecture include Hangzhou's Olympic Sports Center by NBBJ Architects, as well as the Qatar Integrated Railway Project by UNStudio [25], with both design studios producing generations of the buildings using parametric programs and modified design parameters.

When examining the advantages of parametric design, structural engineers in recent years have sought to bring out the potentials of the process. Optimisation is one of the key areas of interest within the context of parametric design, and the easy testability of parametric models through adjustment of configurations is also highly intriguing. With the creation of a parametric model, individuals will not only be able to check different configurations within the changeable parameters, but also be able to directly conduct structural analysis on the model. This practicality is especially applicable to identifying the effect varying parameters have on the performance of certain structures, and reduces the costs that would be obvious in the practice of using separate models for structural analyses. Furthermore, with the relegation of labor-intensive work to the algorithm, the influence of human error would be lessened, as would the labour costs. Coming in altogether, these factors would result, overall, in a higher design efficiency and more efficacious design process [25].

### B. Objective & Motivation

The primary objective of this project is to investigate the feasibility of devising a pathfinding algorithm with the capacity to autonomously generate functional accessibility ramps according to publicised regulations, all while affording users the ability to customise characteristics within a set conformation to regulations and standards.

## II. METHODOLOGY

### A. Software

Using Rhino 7 as the software base of the program, the corresponding parametric design plug-in utilised is Grasshopper. Rhino 7 is a 3D computer-aided design software developed by McNeel & Associates which primarily bases its geometry on NURBS modelling [25]. Here, the methodology proposed divides the process of ramp design into six steps, completing each using parametric design components in Grasshopper and Python code for data processing between each component.

The use of third-party software, here, primarily acts as an imported Python library for streamlining the process of converting the three-dimensional environment into usable data, and vice-versa.

### B. Code Structure

The division of the primary algorithm is partitioned into five main steps: Input, Processing, Optimisation, Actualisation, and Display. Thorough outlining and description of each step is presented below, and a complete visual representation of the code can be observed in Figure 1.

#### 1. Input

Initiating with manual data input, the algorithm begins with the specification of start and end points within a 3D model, along with the establishment of the model boundary. The distance between these points are then processed and utilised to construct a three-dimensional matrix, where the z-value for height signifies vertical disparities between them. This matrix is then used for subsequent processes.

Parametric data customisation then offers the flexibility to adapt the algorithm according to specific requirements, such as path width, slope, thickness, inter-path distance, railing height, and more.

#### 2. Processing

Firstly, for basic processing, the A* pathfinding algorithm is employed to identify the most optimised, or shortest, path between the start and end points on a flat plain. For the sake of ensuring compliance with legislative regulations and the desired slope, the algorithm iteratively readjusts the slope-height of the path. This, then, accommodates for the adaptability needed to meet country-specific requirements.

Once the path is established, height processing begins as height elements are integrated into the trajectory. Incorporating these elements allow for the path to navigate obstacles and varying terrain heights. In the case of multi-layered paths, the algorithm is capable of dynamically transitioning to modified functions to ensure precise path realisation.

#### 3. Optimisation

For the process of path slope optimisation, the algorithm operates within a preset or user-defined range to iteratively re-execute itself with varying desired path slopes if permitted. This process generates a range of paths, from which the optimal version is selected according to adherence to input slope and path length. This approach then guarantees the final path chosen as the most optimised version for the specific site and chosen parameters

#### 4. Actualisation

When it comes to ramp actualisation, by leveraging parametric design elements within the Grasshopper plugin, the algorithm extends and extrudes the path onto a base surface. Then, based on user-defined input parameters and material choices, railings and structural supports are incorporated.

By accommodating for a diverse range of practical and visual requirements, the resulting design is functional and visually-pleasing.

#### 5. Display

Finally, the display of the ramp involves rendering various components based on the inputted material properties. Conclusively, this process provides a clearer visualisation of the ramp's anticipated appearance following construction for consideration.

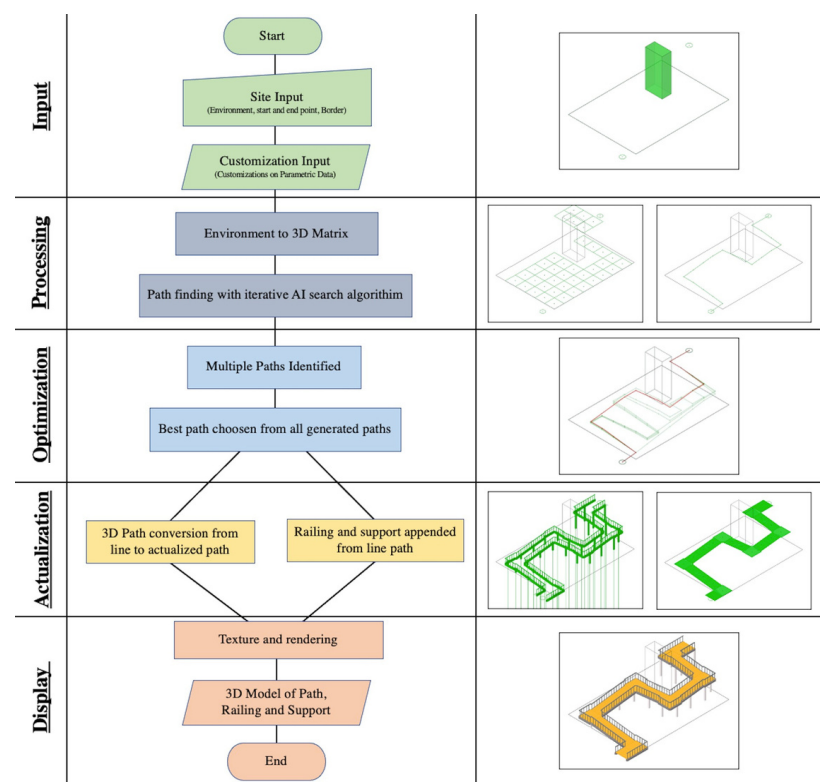

*Fig. 1.* Flow diagram and visual schematics illustrating the five key processing stages.

For technical reference and understanding, a table listing possible input parameters for the generative algorithm are provided in Table 1.

| Sections | Parameters | Description |
|---|---|---|
| **Path** | Slope | Controls the ratio between rise and run of the path |
| | Thickness | Controls the thickness of the path from 0 to inf |
| | Inter-path distance | Controls space between distance between the two rows of path |
| | Height | Controls the empty space above each section of the path |
| | Level landings | Controls the landings is automatic or manually levels based on the position of the starting and ending point |
| | Type | Control if the path is "Curve" or "Straight" |
| **Railing** | Height | Controls the height of railings |
| | Thickness | Controls the thickness of railing |
| | Density | Controls distance between two railings posts |
| | Type | Controls type of railing between "single square railing", "single rounded railing", or "double rounded railing" (Fig.3 for reference) |
| **Supports** | Thickness | Controls the thickness of the support columns structure underneath the ramp |
| | Density | Controls the number of the support structure underneath the ramps |
| **Materials** | Path | Controls the rendered of the path based on customizable materials available in Rhino 7 |
| | Railing | Controls the rendered of the railing based on customizable materials available in Rhino 7 |
| | Support | Controls the rendered of the support based on customizable materials available in Rhino 7 |

*Table 1: Input parameters for the generative algorithm*

**C. Rationale**

With path inputs mainly centred around published requirements of the ADA, the specific inputs, on the other hand, allow for the efficient altering of the algorithm to fit the specific requirements of the user. This catering to exact requirements is what allows for the algorithm to go beyond simple accessibility ramp generation and apply to a greater scope of needs. Both the railing and support inputs, like the path inputs, are based on the ADA's legislation, and similarly provide for alterations requested by the user to cater their final generation to their exact, personal specifications.

**D. User Interface**

For the algorithm to operate, the user simply needs to provide it with an environment boundary, a pair of start and end points, and, optionally, any relevant obstacles in the boundary curve. The remainder of the parameters are preset to adhere to ADA regulations, while they are capable of being adjusted based on user preference.

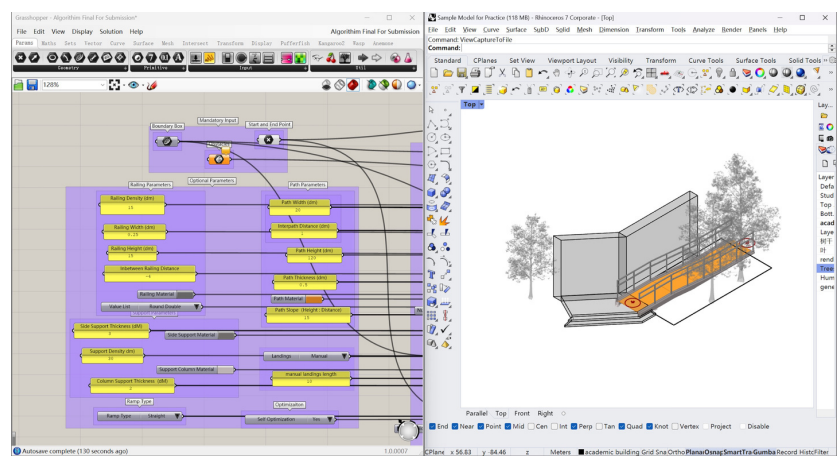

*Fig. 2.* Current user interface of algorithm

### III. CONCLUSION

**A. Verification & Results**

*1. Verification by Simulation*

One of the most optimal tests to analyse the success of the algorithm is to run a set of simulations, each aiming to test different aspects of the algorithm. A total of sixty (60) simulations were run, examining the algorithm's ability to compute different input parameters as provided in Table 1. The results of the simulation were each graded with a score from one (1) to four (4) based on the algorithm's ability to respond with feedback on feasibility, analyse the environment, generate a ramp path, and correctly construct said ramp path. Additionally, to test the efficiency of the algorithm, a timer was set recording the total time taken for the generation of sixty (60) simulations.

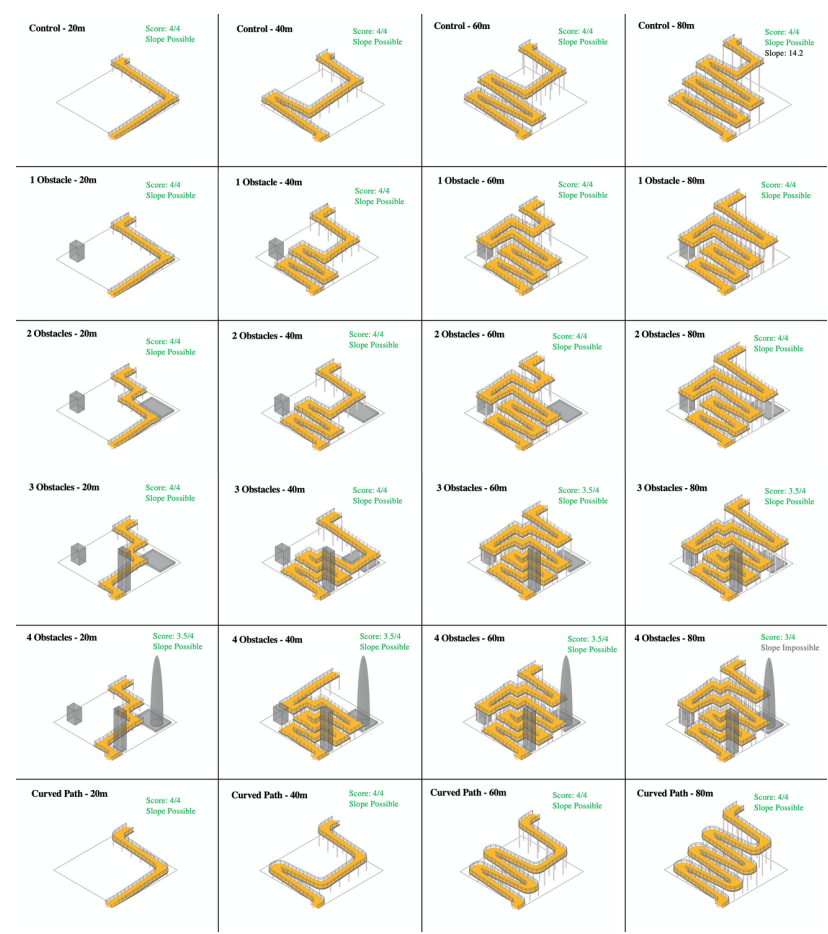

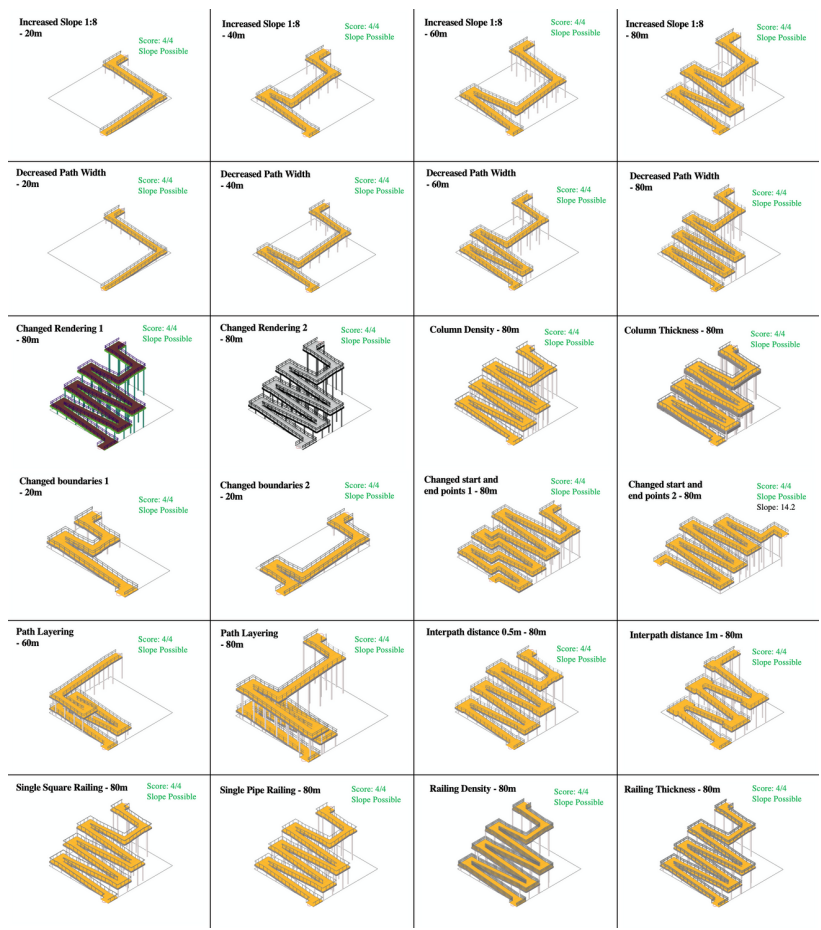

*Fig. 3.* *Simulations run as part of verification*

Based on the results of the simulation, the algorithm can be concluded to function as intended. Every input parameter tested, ranging from path slope to texture rendering, was successfully processed. The only minor issue which arose was when an excessive number of obstacles were introduced in the plane with a high end-height being set. As of now, this issue is not of high concern as the algorithm is unlikely to encounter such a situation in realistic conditions.

In regards to efficiency, all sixty simulations were generated in a total time of around two hours, meaning each ramp generation required around two minutes. In comparison to accessibility ramp design by hand, there is a significant streamlining of the design process through utilising the final algorithm. However, it is relevant to mention here that a few of the simulations were extremely similar to others, meaning that the average time taken per generation is expected to be longer if fulfilling sets of unique requirements.

*2. Implementative Testing*

As further verification, three separate, pre-existing environments containing accessibility ramps were referenced to implement the algorithm on. To conduct these trials accurately, basic measurements were taken of the pre-existing environments, with the ramp removed. In its place, the algorithm would then generate its own accessibility ramp according to the environmental conditions, which would thus be analysed for general feasibility and efficacy.

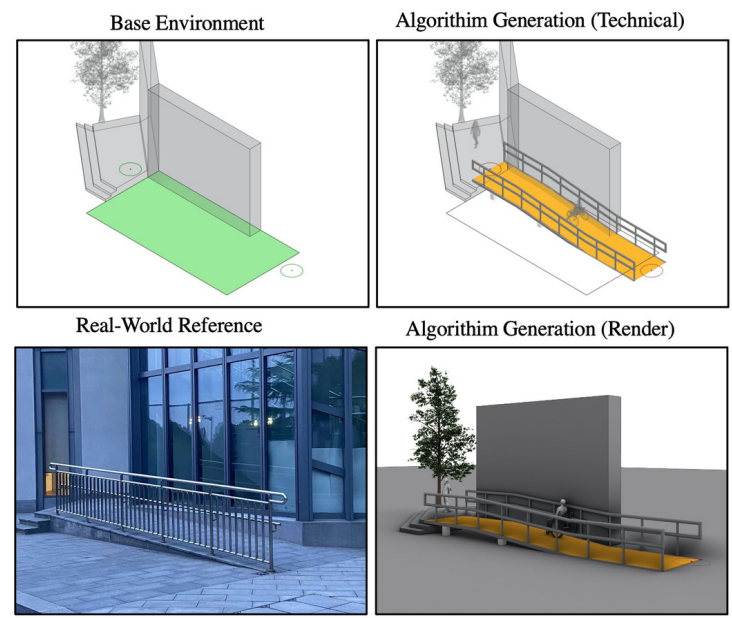

*Fig. 4.* *Comparison of generation-real world for Trial 1*

In this first trial, a simple ramp environment was isolated with a total rose of forty (40) centimetres (cm) and a larger, flat area open to work with. Start and end points are denoted with the circle featuring an 'X' symbol in the centre. The generation from the algorithm is largely similar to the pre-existing ramp, with the minor difference being the extended level landing at the start and end points of the ramp.

Overall, the algorithm was able to successfully generate a ramp which adheres to ADA standards while being optimised and closely resembling the real-world ramps designed by professionals.

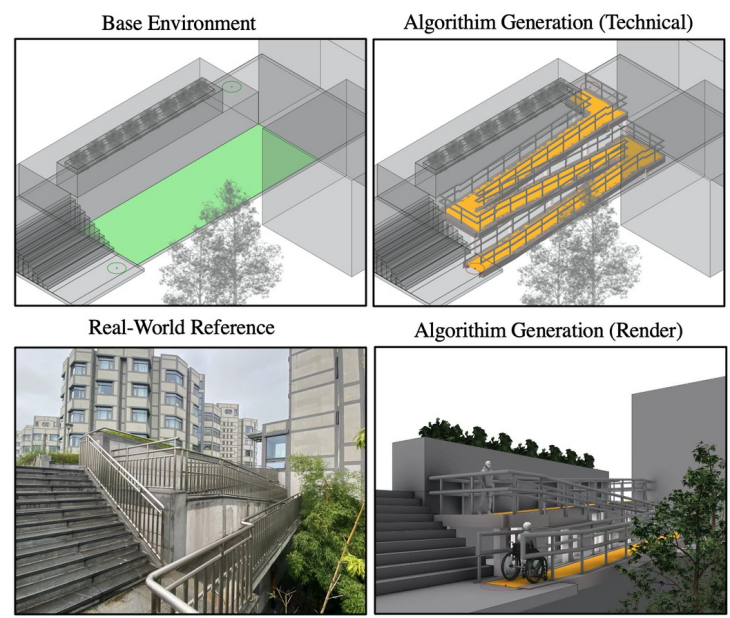

*Fig. 5.* *Comparison of generation-real world for Trial 2*

For the second trial, a more complex ramp environment was located and then isolated. With a rise of two (2) metres (m) and a constrained flat area to work with. Here, the generation from the algorithm was near-identical to the pre-existing ramp, with minor differences being that the end point for the real-world ramp being extended to include additional entrances.

In spite of the more challenging environment, the algorithm was still able to successfully generate an optimised ramp following all listed ADA regulations while being similar, once again, to professionally-designed ramps.

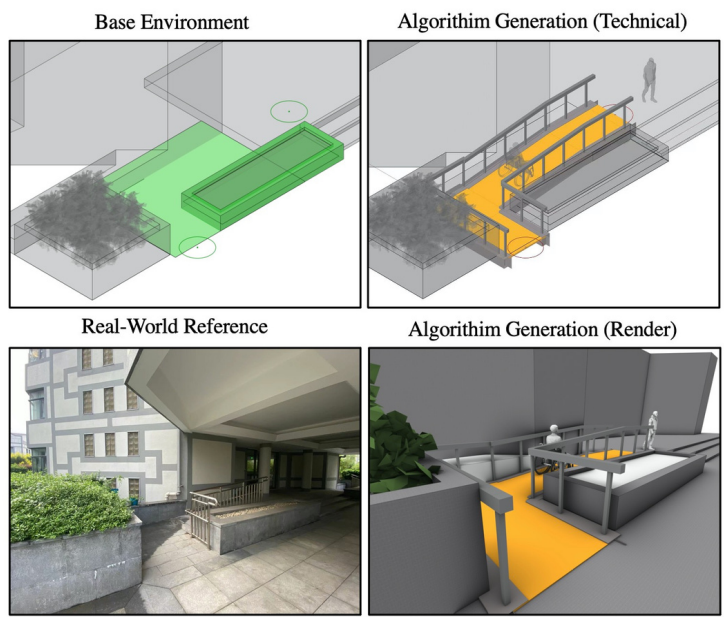

*Fig. 6. Comparison of generation-real world for Trial 3*

In the third and final trial, a ramp environment with a rise of thirty (30) centimetres (cm) and extremely constrained flat area was isolated. The generation here continues to resemble the pre-existing ramp in the real-world location, with the minor difference being the inclusion of railings on both sides of the ramp from the algorithm.

Like before, the algorithm's generation follows ADA requirements, is optimised, and resembles the professional-designed, real-world ramp.

**B. Final Rendering & Further Application**

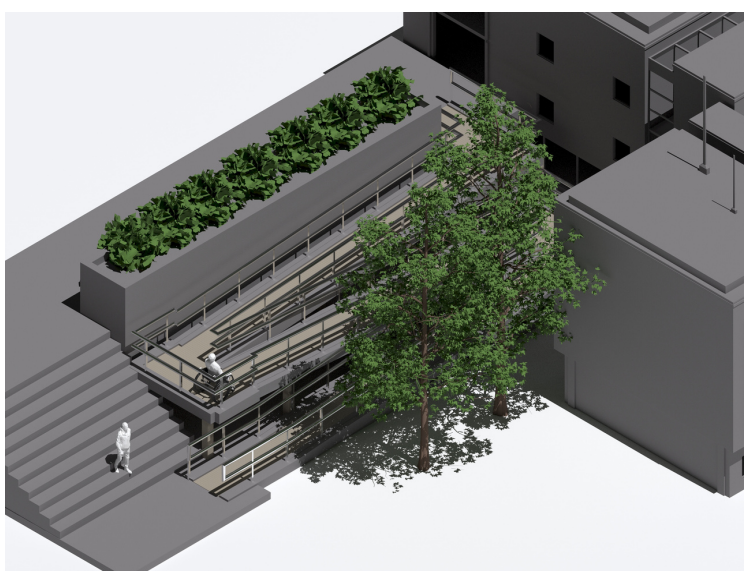

*Fig. 7. Sample rendering of algorithm generation*

To conclude from the simulations, there is no issue with the function of the algorithm for the majority of possible cases, considering its ability to generate ramps near-identical to professionally-designed real-world ramps when provided with relevant environments. As of now, the primary concern with the algorithm's performance is its ability to create circular paths. With the simulations run, it was unable to develop any functioning, working paths for the circular simulations aside from the first one.

Following the completion of the generative process, the final ramp is capable of being directly exported in 3D modelling softwares, allowing for it to be easily rendered and enabling further modification, as well as possibly being exported as a construction blueprint. For example, figure 7 shows a rendering of a ramp generated within a mock environment then exported into Rhino 7 for rendering. As a supplement, it is also feasible to export the 3D model as an STL, DAE, or SDS file, as well as a variety of other formats, allowing the model to be easily transferred amongst other commonly-used city planning and modelling softwares for inter-software exchanges.

**D. Concluding Statements**

In this paper, we propose using parametric design as a suitable medium and solution to streamline the overall design process in implementing accessibility infrastructure. Our final algorithm is able to effectively generate both accurate, acceptable, and optimised accessibility ramps, adhering to all currently available specifications. While being produced in a shorter timeframe than human designers, the generated ramps still resemble the real-life ramps designed and constructed by professionals. For future reference, the ability to generate additional types of accessibility infrastructure and ability to incorporate the algorithm into other softwares as a component or library would increase its ease of use.

## IV. CONTRIBUTIONS

This paper originated from constructive discontent with the lack of availability of accessibility ramps in large cities and is an attempt to remedy this issue. The research background for this paper is mainly in the field of parametric design, python coding, accessibility infrastructure requirements, and AI search algorithms.

*Candidates:*

**Antonio Li:** Methodology, Diagrams, Software, Conceptualization, Validation, Writing – Original Draft

**Leila Yi:** Writing - Review and Editing, Supervision, Research, Resources

*Instructor:*
**Brandon Yeo** – the instructor – is a teacher and advisor at UWC CSC the school which both participants attend. He mainly contributed towards giving advice and feedback in the process of writing the paper and the algorithm. He did not receive any financial compensation for participating in the writing of this paper.